# Transmission singularities in resonant electron tunneling through double complex potential barrier

Nikola Opačak [1(a)], Vitomir Milanović [1(b)] and Jelena Radovanović [1(c)]

[1] School of Electrical Engineering, University of Belgrade - Bulevar Kralja Aleksandra 73, 11120 Belgrade, Serbia

**Abstract** – Tunneling of electrons through a barrier with complex potential is investigated. We focus on two cases, symmetric double rectangular barrier and double delta potential barrier, and give expressions for resonant transmission probability for both cases. Expressions for reflection amplitude and absorption are also obtained in the case of delta potential. It will be shown that for given dimensions of the potential barrier and the real part of the potential, resonant transmission probability approaches infinity for only one positive value of the imaginary part of the potential.

**Introduction.** – Tunneling through a potential barrier is one of the most studied problems in quantum mechanics. This topic includes the case of tunneling where there is a reduction of incident flux due to absorption in the barrier medium. Just as the problem of absorption of electromagnetic waves in a medium could be analyzed in terms of complex refractive index of the medium, it was found that problem of absorption of incident flux could be simulated in terms of complex potential. In quantum mechanics the use of complex potentials was introduced to describe the processes of absorption and generation of electrons in the barrier region [1]. The case of an absorptive material corresponds to the potential with negative imaginary part, and conversely, the case of negative absorption, when we have gain instead of reduction of incident flux, corresponds to potential with positive imaginary part. This model has been successfully used to account for a number of effects: total absorption in barrier or/and well potentials [2], identification of resonances in scattering theory [3], dissipative transmission processes in atomic wires using the array of complex $\delta$ potentials [4], fusion of two heavy nuclei within the concept transmission across absorptive fusion barrier [5]. An extensive review on complex potentials and absorption is given in [6-7].

In this paper we study the tunneling of electrons through two successive potential barriers, to be more specific, we first consider the case of a symmetric double rectangular barrier and subsequently of a double delta potential barrier [8-10]. We focus our analysis on barriers with complex potentials while providing a brief overview of the case with real potential. References [11-13] study transmission through different potentials that could provide some insight into the forthcoming problem. Metastable states exist in the well region between the barriers [14]. The resonant tunneling occurs when the incident electron energy is equal to the metastable resonant state energy in the well.

We obtain the expression for transmission amplitude at resonance condition in case of double barriers both for complex and real potential. It will be shown that for potentials which have negative imaginary part, the resonant transmission probability is always below unity. For strictly real potentials, resonant transmission probability is exactly equal to 1. The situation where we have potentials with positive imaginary part is more difficult to analyze. What seems to be very counter intuitive and interesting, for positive imaginary part of the potential, the resonant transmission probability increases its value until, for one particular value of the imaginary part of the potential, it approaches infinity. After that point, as we further increase the imaginary part of the potential, the resonant transmission amplitude decreases monotonically, and approaches the value zero. This effect, the phenomenon of infinite transmission, will be the main subject of investigation in this paper. Although the effect seems rather strange, similar behavior of the transmission coefficient has been noticed before, so this is not an isolated case. There are other mechanisms which are responsible for incompleteness of the eigenfunctions of the Hamiltonian. They are associated with the presence of exceptional points [15] and spectral singularities [16]. Spectral singularities are among generic mathematical features of complex scattering potentials and are associated with the energies of scattering states having infinite reflection and transmission coefficients. Because this is a characteristic property of resonance states, spectral singularities correspond to the resonance states having a real energy and zero width and have application in designing a waveguide that functions as a resonator, see [17-18]. An example of such potential which supports spectral singularities would be a complex PT-symmetric barrier potential with a corresponding non-

[(a)] E-mail: opacaknn@gmail.com (corresponding author)
[(b)] E-mail: milanovic@etf.bg.ac.rs
[(c)] E-mail: radovanovic@etf.bg.ac.rs

.

Hermitian Hamiltonian [19-21]. Additionally, we will show that the occurrence of infinite transmission exists both for double potential barriers and double potential wells. In other words, for every value of the real part of the potential (positive or negative), there exists a positive value of the imaginary part of the potential, so that the transmission approaches infinity at a resonance.

**Theoretical consideration.** – We investigate a one-dimensional symmetric double barrier structure with complex potential in the following form:

$$U_{(x)} = \begin{cases} 0, x \in \left(-\infty, -\frac{w}{2}-b\right) \cup \left(-\frac{w}{2}, \frac{w}{2}\right) \cup \left(\frac{w}{2}+b, +\infty\right) \\ U_0, x \in \left(-\frac{w}{2}-b, -\frac{w}{2}\right) \cup \left(\frac{w}{2}, \frac{w}{2}+b\right) \end{cases}, \quad (1)$$

where $b$ is the width of each barrier, $w$ is the width of the potential well between them and $U_0 = U_{0R} + iU_{0I}$.

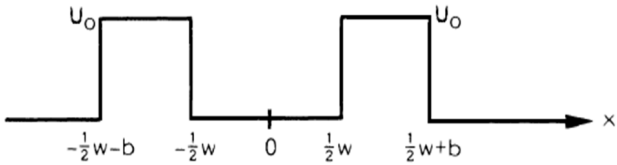

Fig. 1: A symmetric rectangular double barrier. The height of the barriers is $U_0$, their width is $b$, and the distance between them is $w$.

The complex transmission amplitude is found to be (similarly as in [10]):

$$T(k) = e^{-i2kb}/D(k), \quad (2)$$
$$D(k) = \cosh^2(\kappa b) + \frac{1}{4}\sinh^2(\kappa b)[\sigma^2 \cos(2kw) - \delta^2] +$$
$$i \sinh(\kappa b)\left[\delta\cosh(\kappa b) + \frac{1}{4}\sigma^2 \sinh(\kappa b)\sin(2kw)\right],$$

where:

$$k = \sqrt{\frac{2m}{\hbar^2}E}, \kappa = \sqrt{\frac{2m}{\hbar^2}(U_0 - E)}, \delta = \frac{\kappa^2 - k^2}{\kappa k}, \sigma^2 = \delta^2 + 4 \quad (3)$$

On introducing the functions:

$$U = \cosh^2(\kappa b) - \frac{1}{4}\delta^2 \sinh^2(\kappa b),$$
$$V = \delta\cosh(\kappa b)\sinh(\kappa b), W = \frac{1}{4}\sigma^2 \sinh^2(\kappa b), \quad (4)$$

one can conclude that they are connected by relation:

$$U^2 + V^2 = (1 + W)^2. \quad (5)$$

Next, by using the functions (4), the expression for $D(k)$ can be written in the form:

$$D(k) = U + W \cos(2kw) + i[V + W \sin(2kw)],$$

$$D(k) = [U_R + W_R \cos(2kw) - V_I - W_I \sin(2kw)] +$$
$$i[U_I + W_I \cos(2kw) + V_R + W_R \sin(2kw)], \quad (6)$$

where $U_R, V_R, W_R$ are the real parts of $U, V, W$, respectively, while $U_I, V_I, W_I$ are the corresponding imaginary parts.

We are interested in obtaining the expression for the transmission probability of the resonant state. The resonance condition can be expressed in various ways, out of which, we will use the following relation that gives extremal condition, which can be applied to asymmetric cases as well:

$$\frac{d}{dw}|D(k)|^2 = 0. \quad (7)$$

From (7), after finding the expression for $|D(k)|^2$ and differentiating it with respect to $w$, we obtain the equation which is valid in case of transmission probability extrema, both for minima and maxima of the transmission:

$$\text{tg}(2kw) = \frac{-U_R W_I + V_I W_I + U_I W_R + V_R W_R}{U_R W_R - V_I W_R + U_I W_I + V_R W_I} = \frac{M}{N}. \quad (8)$$

We have introduced functions M and N for the purpose of writing the previous relation more compactly. Relation (8) can be transformed to a much simpler equation $\text{tg}(2kw) = V_R/V_I$, when the potential is strictly real ($U_{0I} = 0$).

The resonances correspond to those solutions of (8) at which:

$$\sin(2kw) = -\frac{M}{\sqrt{M^2+N^2}}. \quad (9)$$

After combining relations (9) and (6), after some calculations, we obtain the final form of $D(k)$ at resonance:

$$D(k)_{res} = (U + iV)\left[1 - \sqrt{\frac{W_R^2 + W_I^2}{(U_R - V_I)^2 + (V_R + U_I)^2}}\right]. \quad (10)$$

It is important to point out that the expression in square brackets can reach the zero value for complex potentials. We will show that this effect occurs only for one particular positive value of imaginary part of the potential $U_{0I}$, if the $U_{0R}$ is already given. In such a case, $D(k)_{res} = 0$, and the transmission probability for resonance condition $|T_{res}|^2$ approaches infinity. We will investigate this effect in more detail later in the paper, in the case of delta potential barriers, because the corresponding expressions are simpler, however, similar procedure and calculations can be made in the case of rectangular barriers, with the same result.

If we, again, consider strictly real potential barriers ($U_{0I} = 0$), and take (5) and (4) into consideration, expression (10) transforms into a much simpler form [10]:

$$D(k)_{res} = \frac{U+iV}{1+W} = \frac{1-\frac{1}{4}\delta^2 \tanh^2(\kappa b) + i\delta \tanh(\kappa b)}{1+\frac{1}{4}\delta^2 \tanh^2(\kappa b)}, \quad (11)$$

which clearly shows that $|T_{res}|^2 = |D_{res}|^{-2} = 1$, for real potential barriers, as expected. This is not the case when we have absorption or generation of electrons in the barrier region, described by complex potentials. For potentials with negative imaginary part (absorption), the resonant transmission probability is $|T_{res}|^2 < 1$. In case of potentials with positive imaginary part, the behavior is more complicated for analysis.

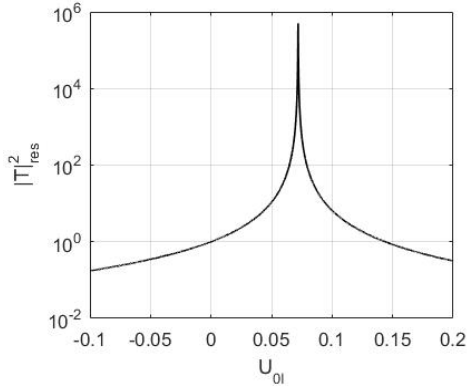

Fig. 2: Transmission probability at the resonance condition as a function of the imaginary part of potential, $U_{0I}$ (in units of meV). The maximum value is obtained for $U_{0Ipeak} = 71.917$ μeV, and for resonant energy $E_0 = 0.1194\ eV$. Barrier widths are $b = 5$ nm, and well width $w = 5$ nm. Effective mass $m = 0.067\ m_0$, where $m_0$ is electron rest mass. The real part of the potential is taken to be $U_{0R} = 0.7$ eV.

Figure 2. depicts the dependence of $|T_{res}|^2$ on the imaginary part of the potential $U_{0I}$. It is seen that for real potential barriers (when $U_{0I} = 0$) the resonant transmission probability is $|T_{res}|^2 = 1$. It can be also observed, that for potentials with negative imaginary part ($U_{0I} < 0$), it holds $|T_{res}|^2 < 1$, and generally, the resonant transmission amplitude approaches zero value with the decrease of $U_{0I}$. This is expected, since potentials with negative imaginary parts describe absorptive materials. However, for $U_{0I} > 0$, the transmission probability shows very interesting behavior. It increases, has a peak with very high value (which will be shown to approach infinity), and later decreases monotonically and approaches the value of zero. For parameters used in example depicted in fig.1., value of $U_{0I}$ for which the transmission peak occurs is $U_{0Ipeak} = 71.917$ μeV.

**Double Delta potential** – We will now consider that barrier height $U_0 \to \infty$, and at the same time, the barrier width $b \to 0$ while we keep the product $U_0 \cdot b = V_0$ constant. The width of the well between the barriers $w$ is unchanged. In other words, we will investigate the symmetrical centralized double delta potential structure of the following form:

$$U(x) = V_0 \left( \delta\left(x + \frac{w}{2}\right) + \delta\left(x - \frac{w}{2}\right) \right). \tag{12}$$

Considering the delta potential, we can make the following approximations:

$$\cosh(\kappa b) \approx 1,\ \sinh(\kappa b) \approx \kappa b,\ \sigma \approx \delta \approx \sqrt{U_0/E}. \tag{13}$$

When we combine these approximations with the expression (2) for $D(k)$ it follows:

$$D(k) = 1 + \frac{m}{2\hbar^2}\frac{V_0^2}{E}[\cos(2kw) - 1] + i\sqrt{\frac{2m}{\hbar^2}}\frac{V_0}{\sqrt{E}} + i\frac{m}{2\hbar^2}\frac{V_0^2}{E}\sin(2kw). \tag{14}$$

If we introduce:

$$a = \frac{m}{2\hbar^2 E} \tag{15}$$

then $D(k)$ can be expressed in the following manner:

$$D(k) = 1 - aV_0^2[1 - e^{i2kw}] + i\,2\sqrt{a}V_0. \tag{16}$$

From (4), $U, V, W$ can be written in the following form:

$$\begin{aligned}&U = U_R + i\,U_I,\ U_R \approx 1 - a(V_{0R}^2 - V_{0I}^2),\ U_I \approx -2aV_{0R}V_{0I},\\&V = V_R + iV_I,\ V_R \approx 2\sqrt{a}V_{0R},\ V_I \approx 2\sqrt{a}V_{0I},\\&W = W_R + iW_I,\ W_R \approx a(V_{0R}^2 - V_{0I}^2),\ W_I \approx 2aV_{0R}V_{0I}.\end{aligned} \tag{17}$$

The extremal condition is again (7), so after finding $|D(k)|^2$ and differentiating with respect to $w$ (similar procedure as with rectangular barriers), one gets an eq. analogous eq. (8):

$$\text{tg}(2kw) = \frac{M}{N}, \tag{18}$$

where:

$$M = 2a^2 V_{0R}V_{0I}(V_{0R}^2 - V_{0I}^2) - 2aV_{0R}V_{0I} + 4a^{3/2}V_{0R}V_{0I}^2 - 2a^2 V_{0R}V_{0I}(V_{0R}^2 - V_{0I}^2) + 2a^{3/2}V_{0R}(V_{0R}^2 - V_{0I}^2),$$

$$N = 4a^{3/2}V_{0I}V_{0R}^2 + a(V_{0R}^2 - V_{0I}^2) - a^2(V_{0R}^2 - V_{0I}^2)^2 - 2a^{3/2}V_{0I}(V_{0R}^2 - V_{0I}^2) - 4(aV_{0R}V_{0I})^2. \tag{19}$$

In the last eq. (19), $V_{0R}$ and $V_{0I}$ are real and imaginary parts of $V_0$, respectively.. Equations (6) and (19) yield:

$$D(k) = U + iV + W\sin(2kw)\frac{N + iM}{M}. \tag{20}$$

From eq. (18) one can arrive at resonance condition in the same form as relation (9) (but with suitable expressions for $M$ and $N$ given by relation (19)).

Combining relation (20) with the resonance condition (9) (similar procedure as with rectangular barriers), and after inserting relation (19) for $M$ and $N$, one obtains the expression for $D(k)$ at resonance:

$$D(k)_{res} = \left(1 + i\sqrt{a}V_0\right)^2 \left[1 - \frac{a(V_{0R}^2 + V_{0I}^2)}{\sqrt{\left(1 - a(V_{0R}^2 - V_{0I}^2) - 2\sqrt{a}V_{0I}\right)^2 + \left(2\sqrt{a}V_{0R} - 2aV_{0R}V_{0I}\right)^2}}\right]. \quad (21)$$

Equation (21) could also be derived from equation (10) after combining it with eq. (17).

Figure 3. shows very similar characteristics as figure 2. Again, we have the phenomenon of very high peak of the transmission probability for the resonant state, which occurs for one particular positive vale of $V_{0I}$ (the peak occurs when $V_{0I} = V_{0Ipeak}$). It will be shown that the value of this peak approaches infinity (as in the case of rectangular potential barriers). For values of the imaginary part of the potential larger than $V_{0Ipeak}$ ($V_{0I} > V_{0Ipeak}$), the resonant transmission probability monotonically decreases and approaches zero with the growth of $V_{0I}$. As for the region of negative imaginary part of potential ($V_{0I} < 0$), the resonant transmission probability is $|T_{res}|^2 < 1$, and also approaches zero with the decrease of $V_{0I}$. When we have a strictly real potential ($V_{0I} = 0$), it can be observed that the resonant transmission probability is $|T_{res}|^2 = 1$. The distance between two delta barriers is $w = 3nm$, effective mass $m = 0.067\, m_0$ and real part of $V_0$ is $V_{0R} = 2.3\, nm \cdot eV$. This parameters will be used in all further calculations.

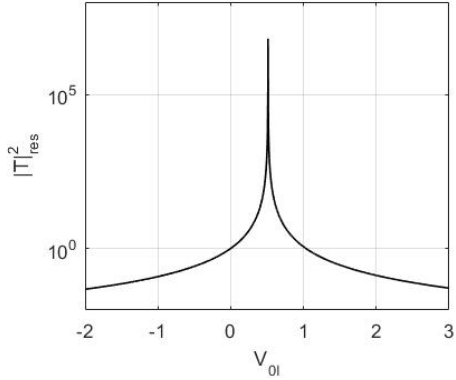

Fig. 3: Transmission probability for the double delta barrier structure at resonance as a function of imaginary part of $V_0$ (in units of $nm \cdot eV$). The maximum value is obtained for $V_{0Ipeak} = 0.5131\, nm \cdot eV$, and for resonant energy $E_0 = 0.4622\, eV$.

We will now investigate the maximum of the transmission probability $|T_{res}|^2$ and prove that its value is infinitely high. We can see from equation (21) that $\text{Re}\{D_{res}\}$ and $\text{Im}\{D_{res}\}$ both have a mutual factor. In order for transmission to be infinite, the condition $D_{res} = 0$ must be satisfied, and that implies that the mutual factor must be equal to zero. If that is the case, then the following relation can be obtained from (21):

$$a^2(V_{0R}^2 - V_{0I}^2)^2 = \left(1 - a(V_{0R}^2 - V_{0I}^2) - 2\sqrt{a}V_{0I}\right)^2 + \left(2\sqrt{a}V_{0R} - 2aV_{0R}V_{0I}\right)^2, \quad (22)$$

which yields a nonlinear equation in terms of $V_{0I}$:

$$4a^{3/2}V_{0I}^3 - 6aV_{0I}^2 + 4a^{1/2}(1 + aV_{0R}^2)V_{0I} - 2aV_{0R}^2 - 1 = 0. \quad (23)$$

It can be shown through numerical analysis that this equation has one real root (fig.4.) for values of $V_{0I}$ that are relevant, thus proving that for only one particular value of $V_{0I}$, equal to the value of the root of the equation (23), $D_{res}$ is equal to zero. When that is the case, the value of resonant transmission probability approaches infinity. Analogous procedure can be conducted to prove the same phenomenon of infinite transmission in the case of rectangular barriers (fig.2).

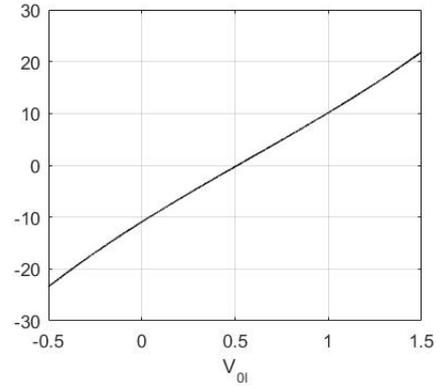

Fig. 4: Value of the expression on the left hand side of equation (26) as a function of $V_{0I}$ (in units of $nm \cdot eV$). The equation is satisfied for $V_{0I} = 0.5131\, nm \cdot eV$ (the same value of $V_{0I}$ where resonant transmission maximum occurs in Fig.3.).

We can also state that along with transmission, the reflection and absorption probability approach infinity when $V_{0I} = V_{0Ipeak}$. At the same time, the quasi-unitary condition, that states that the sum of transmission, reflection and absorption probabilities is equal to unity [1], is satisfied for all values of potential except for $V_{0I} = V_{0Ipeak}$, where we have singularity.

The expression for reflection amplitude $R$ for our case, which is symmetrical centralized double delta barrier is well-known and easy to get hold of considering the condition for continuity of wave functions and known condition for the first derivative when we have delta potential:

$$R = \alpha \frac{(\alpha - 2i)e^{-ikw} - (\alpha + 2i)e^{ikw}}{\alpha^2(e^{i2kw} - 1) + 4i\alpha + 4}, \quad (24)$$

where $\alpha = \sqrt{\frac{2m}{\hbar^2 E}} V_0$.

For the sake of simplifying the upcoming calculations we will translate the given potential (16) for $w/2$, so the new potential is: $U(x) = V_0\big(\delta(x) + \delta(x - w)\big)$. As a result, we obtain the expression for reflection amplitude $R'$ for the new potential, considering a wave incident from the left:

$$R' = Re^{ikw}. \tag{25}$$

The expression for absorption $A$ is [1]:

$$A = -\frac{2m}{\hbar^2 k}\int_{-\varepsilon}^{w+\varepsilon} U_i |\psi_{(x)}|^2 \, dx = -\frac{2m}{\hbar^2 k} V_{0i}\left(|\psi_{(x=0)}|^2 + |\psi_{(x=w)}|^2\right), \tag{26}$$

where $U_i$ is the imaginary part of the given potential $U(x)$, and $\varepsilon \to 0$.

For our case, considering particles incident from the left of the barrier system, the wave function is of the following form:

$$\psi(x) = \begin{cases} e^{ikx} + R'e^{-ikx}, & x \leq 0 \\ Te^{ikx}, & x \geq w \end{cases}. \tag{27}$$

The interval $x \in (0, w)$ is not listed because it is not significant for our analysis. Now, after combining eqs. (26) and (27), and after some calculations, we obtain the expression which links absorption $A$ and reflection amplitude $R$:

$$A = \frac{2}{1 - \frac{\hbar^2 k}{2mV_{0i}}}\left(1 + Re\{Re^{ikw}\}\right). \tag{28}$$

The resonant reflection and absorption probabilities as functions of $V_{0I}$ are depicted on figs. 5. and 6. respectively.

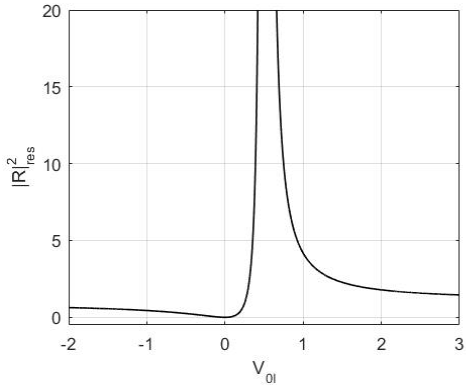

Fig. 5: Reflection probability for the double delta barrier structure at resonance as a function of imaginary part of $V_0$ (in units of $nm \cdot eV$).

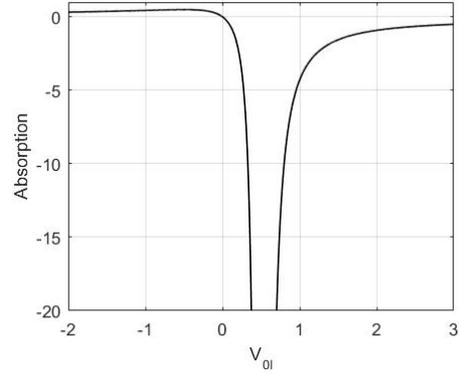

Fig. 6: Absorption probability for the double delta barrier structure at resonance as a function of imaginary part of $V_0$ (in units of $nm \cdot eV$).

In fig.5. it can be clearly seen that for strictly real potential the reflection amplitude vanishes for resonant tunneling, as is expected. When $V_{0i} = V_{0Ipeak}$, reflection approaches infinity, confirming our previous statement. For values of $V_{0i}$ approaching $\pm\infty$, reflection probability approaches the value of zero, which effectively means that the incident particle sees infinitely high barrier and reflects upon it, although the real part of the potential is finite. Figure 6. depicts absorption at a resonance. For negative values of $V_{0i}$, the absorption is positive and approaches the value of zero as the $V_{0i}$ decreases further. When $V_{0i}$ is positive, the absorption is negative, which is expected, because in that case we no longer have absorptive medium. For $V_{0i} = V_{0Ipeak}$ the absorption approaches negative infinity, and for strictly real potential it reaches zero.

After showing that $D_{res}$ is equal to zero when $V_{0I} = V_{0Ipeak}$, next logical step would be to show, if possible, that there exists an equation which links $V_{0R}$, $V_{0I}$ and energy of the electron $E$ in that point, where we have transmission singularity. If $D_{res} = 0$ then we can write (16) in the following form:

$$\alpha^2[1 - e^{i2kw}] - i\,4\alpha - 4 = 0. \tag{29}$$

Last expression (29) is a quadratic equation from which we can easily obtain the roots $\alpha_{1/2}$:

$$\alpha_{1/2} = -\frac{\cos(kw) \pm 1}{\sin(kw)} + i. \tag{30}$$

After replacing the expression for $\alpha = \sqrt{2m/\hbar^2 E}\, V_0$ in the last relation (30), we can get a pair of equations which give $V_{0R}$ and $V_{0I}$ as functions of resonant electron energy:

$$V_{0R} = -\sqrt{\frac{\hbar^2 E}{2m}}\frac{\cos(kw) \pm 1}{\sin(kw)}, \tag{31}$$

$$V_{0I} = \sqrt{\frac{\hbar^2 E}{2m}}. \tag{32}$$

In relation (31) value of $V_{0R}$ must be nonnegative in order for potential to form a barrier, so we choose the appropriate sign depending on the value of energy $E$, and barrier parameters (a corresponding procedure takes place if we chose $V_{0R}$ to be negative and form potential wells). It can also be concluded from relation (32) that transmission singularity only occurs if and only if $V_{0I}$ is a positive value, as we have stated before.

After combining (31) and (32) we finally get the expression which links values of $V_{0R}$ and $V_{0I}$ at singular point, where the transmission approaches infinity (for the case of potential barrier, potential wells will be analysed shortly after):

$$V_{0R} = \begin{cases} V_{0I}\text{tg}\left(\frac{mw}{\hbar^2}V_{0I}\right), & \frac{mw}{\hbar^2}V_{0I} \in \left(n\pi, n\pi + \frac{\pi}{2}\right) \\ -V_{0I}\text{ctg}\left(\frac{mw}{\hbar^2}V_{0I}\right), & \frac{mw}{\hbar^2}V_{0I} \in \left(n\pi + \frac{\pi}{2}, n\pi + \pi\right) \end{cases}, \tag{33}$$

where $n = 0,1,2 \ldots$ for positive values of $V_{0I}$.

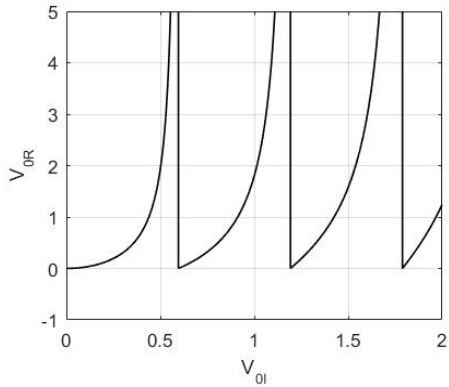

Fig. 7: $V_{0R}$ as a function of $V_{0I}$ (both in units of nm · eV) when the conditions for infinite transmission are satisfied. The distance between two delta barriers is $w = 3$ nm, effective mass $m = 0.067\, m_0$. We have chosen the case of potential barriers where $V_{0R}$ is positive.

Relation (33) is depicted in fig.7. It shows dependence of $V_{0R}$ on $V_{0I}$ at resonance, when the barrier parameters, effective mass and resonant energy are such that $D_{res} = 0$ is satisfied and we have transmission singularity. It can also be observed that that for every value of $V_{0R}$, no matter how big it is, we can find a set of parameters so that the phenomenon of infinite transmission occurs.

If we choose the potential with negative values of $V_{0R}$, we have potential wells instead of barriers and obtain the same relation as (33) only with ranges for values of $mwV_{0I}/\hbar^2$ switched places with one another. In other words, for every value of $V_{0R}$ (negative for potential wells or positive for potential barriers) we can find a corresponding positive value of $V_{0I}$ so that transmission approaches infinity at a resonance. For instance, if we chose $V_{0R} = -2.3$ nm · eV, from $V_{0R} = V_{0I}\text{tg}(mwV_{0I}/\hbar^2)$ we get that corresponding value of imaginary part of the potential is $V_{0I} = 0.71$ nm · eV. We kept the initial distance between the wells $w = 3$ nm.

**Conclusion –** In this paper we have investigated the electron tunneling through two successive complex potential barriers. We have focused on two cases, symmetric double rectangular barrier and a double delta potential barrier, and derived analytical results for the transmission amplitude at resonance condition for both cases.

First, we have considered a double rectangular complex barrier, and have obtained corresponding expression for the resonant transmission amplitude. Through numerical analysis of the acquired relations, with emphasis on potentials with positive imaginary part, we have shown that the transmission amplitude at resonance condition approaches infinity. This effect occurs for only one particular positive value of the imaginary part of the potential, when the real part is fixed. Interestingly, the imaginary part of the potential, for which the transmission singularity occurs, is relatively small compared to the value of the real part of the potential.

Afterwards, we focused on the double delta potential barrier because the relevant expressions are shorter and easier to acquire and analyze. Again, we have shown that for the given real part of the potential, and well width, the same phenomenon of infinite probability occurs for one particular positive value of imaginary part of the potential. We also obtained expressions for reflection amplitude and absorption for this barrier structure, and have shown that they, along with transmission, have singularities for the same values of parameters. Ultimately, a relationship between energy, real and imaginary part of the potential is determined at the point of singularity and it is shown that the occurrence of infinite transmission exists both for double potential barriers and double potential wells as long as the imaginary part of the potential has positive value.

***

This work was supported by the Ministry of Education, Science and Technological Development of the Republic of Serbia under Project No. III 45010.
.